# The onset and growth of the 2018 Martian Global Dust Storm

A. Sánchez-Lavega[1,2], T. del Río-Gaztelurrutia[1,2], J. Hernández-Bernal[1,2], M. Delcroix[3]

1. Dpto. Física Aplicada I, EIB, Universidad País Vasco UPV/EHU, Bilbao, Spain

2. Aula EspaZio Gela, Escuela de Ingeniería de Bilbao, Universidad del País Vasco UPV/EHU, Bilbao, Spain

3. Societé Astronomique de France, Paris, France

Corresponding author: Agustín Sánchez-Lavega (agustin.sanchez@ehu.eus)

**Key Points:**

- May-June 2018 ground-based images show the onset and early evolution of a Martian Global Dust Storm (GDS)

- The outbreak took place at location (North hemisphere) and time (solar longitude 184.9°) unusual for most GDSs

- The expansion involved horizontal velocities in all directions in the range 5-40 ms$^{-1}$

## Abstract

We analyze the onset and initial expansion of the 2018 Martian Global Dust Storm (GDS 2018) using ground-based images in the visual range. This is the first case of a confirmed GDS initiating in the Northern Hemisphere. A dusty area extending about $1.4x10^5$ km$^2$ and centered at latitude +31.7°±1.8° and west longitude 18°±5°W in Acidalia Planitia was captured on 30 and 31 May 2018 (Ls = 184.9°). From 1 to 8 June, daily image series showed the storm expanding southwards along the Acidalia corridor with velocities of 5 ms$^{-1}$, and simultaneously progressing eastwards and westwards with horizontal velocities ranging from 5 to 40 ms$^{-1}$. By 8 June the dust reached latitude -55° and later penetrated in the South polar region, whereas in the North the dust progression stopped at latitude ∼ +46°. We compare the onset and expansion stage of this GDS with the previous confirmed storms.

## 1. Introduction

Mars Global Dust Storms (GDS), also referred as "planet encircling dust storms" and "Global Dust Event", represent unique dynamical phenomena in terrestrial planet atmospheres (see Khare et al., 2017 and Montabone and Forget, 2018a for recent reviews on the phenomenon, and references therein). They are unusual and nowadays unpredictable aperiodic events (in time and location), with the few confirmed cases observed in years 1956, 1971, 1973, 1977A, 1977B, 2001 and 2007 (Martin and Zurek, 1973; Thorpe 1979, 1981; Ryan and Sharman, 1981; Strausberg et al., 2005; Wang and Richardson, 2015; Khare et al., 2017; Montabone and Forget, 2018a, 2018b). A summary of the properties of these GDS is given in Supporting Information Table S1. The onset of all confirmed GDSs occurred from Ls = 185° to 300°, during the "high dust loading season" that begins approximately with the autumn equinox in the northern hemisphere at orbital longitude Ls = 180° and spans about a half Martian year to Ls = 360° (Khare et al., 2017). It should be noted that major storms in MY 27 and 29 started somewhat earlier, at Ls ∼ 140°, but they were not confirmed as GDSs (Montabone and Forget, 2018a, 2018b). The subject of this paper is the analysis of the onset phase and early evolution of a new GDS that initiated on 30 May 2018 (Martian Year MY 34) at West longitude 18°±5°W and latitude +31.7°±1.8° in Acidalia Planitia. Two important features of the 2018 GDS event distinguish it from previous known cases. First, starting at Ls = 185°, it is the earliest in the Martian seasonal cycle together with the 2001 case (Cantor, 2007). Second, even though the Acidalia Planitia is a common place for regional dust storms (Wang and Richardson, 2015), no other GDS has been seen to initiate in the northern hemisphere (Table S1).

Here we report an analysis of ground-based images obtained in the visual range by a large network of amateur astronomers contributing to the "Planetary Virtual Observatory Laboratory" PVOL (Hueso et al. 2018) and the "Association of Lunar and Planetary Observers" ALPO Japan repositories of planetary images and to specialized internet blogs. In Table S2, we provide details on the contributors, their telescopes, and dates of the observations used in this work. The period studied spans from May 27 to June 8, which corresponds to the onset and early expansion stages of the GDS, and we complement it with a qualitative description of the evolution of the GDS up to June 27, when most of the dust encircled the planet. Images showed signs of the initial clearing of the dust by July 2018 (Walker, 2019) but the study of the declining stage of the GDS is left for the future. In this work, we also compare our measurements with available data from studies of the outbreak and onset phases of previous GDSs. Finally, we discuss the retrieved wind speeds and dust area expansion within the framework of the Mars Climate Database (MCD), which provides a statistical summary of the output of the Global Climate Model of the Laboratoire de Météorologie Dynamique (LMD-MGCM) (Forget et al., 1999; Millour et al., 2015). The study also shows the importance of the imaging survey provided by expert amateur astronomers to follow the atmospheric dynamics of Mars (Sánchez-Lavega et al., 2015) and other planets.

## 2. Onset and Initial Expansion of the Global Dust Storm

The analysis of the images, and the procedures followed to measure the location on the Martian disk of different features, are the standard ones used in previous works (Sánchez-Lavega et al., 2015) with a few additional aspects described in the Supplementary Information. Summarizing, we navigated the images (i.e. determined the Martian disk edge and the longitude-latitude grids on the disk) using the WinJUPOS software (2019). This allowed us to obtain the coordinates of any feature on the planet, measure the area covered by the dust as a function of time, and retrieve the motions of the most prominent features at the edges of the dust cloud.

The series of images in Figure 1 shows the onset phase of the GDS 2018. High quality ground-based images of the location of the onset of the storm obtained along May, showed no sign of activity (Fig. 1a), but on May 27 a large yellowish area was captured in the north at ~ 347°W and 56°N (Fig. 1b), presumably corresponding to dust that we consider as a precursor storm. On May 30, a small dusty area is observed southward of the precursor, centered at ~12°W and 35°N corresponding to the Acidalia Planitia (Fig. 1c). It was first imaged at 7:25 U.T. corresponding to Local Mars Solar Time = 11:54 and Ls= 184.2° (MY = 34, sol-of-year 380). This storm occupied an area of $1.4\pm0.3\times10^5$ km$^2$ and we consider it the outbreak area of the GDS2018. On May 31 (sol-of-year 381), the main body of the storm (the region occupied by the dust with the highest optical depth) had expanded considerably to an area of $5.1\pm0.3\times10^5$ km$^2$ (still a local storm according to Cantor, 2007) while its center remained at the same location within the measurement uncertainty (Fig. 1d). Simultaneously a north-east expansion of the dust by the winds took place, covering a larger area but with a lower optical depth, extending up to ~ 335°W and 45°N (Fig. 1d, right arrow). This represents a distance of ~ 1,600 km from the edge of the main body of the storm. On 1 June, two new dusty areas formed westward of the first storm, apparently elongated in the meridional direction (Fig. 1e).

The expansion of the storm from 2 to 5 June took place in the westward and south directions. In the west it formed at least three different conspicuous north-south meridional "branches" that grew sequentially in longitude, between ~ 0° to 60° W, with the winds driving the dust towards the equator (Fig. 1f-j). These meridional branches were triggered west and south from the GDS onset area (Fig. 1 f-h). The westward branch can be seen in Fig.1i and the complete three branches in Fig. 1j. The dust from these branches mixed and the area became homogeneously covered with dust by June 6 (Fig. 1k). Simultaneously, a slower eastward expansion occurred. However, the eastern border of the storm was less contrasted and difficult to track when compared to the west border (Fig. 1l). At the same time a local storm took place on June 6 west of Hellas (302°W and 35°S-50°S, Fig. S1) not interfering with the GDS. According to MARCI/MRO movies (Malin 2018a, 2018b) another local storm developed on June 6-7 (again independently of the GDS) at approximately 55°S and 0°W. The continuous expansion to the South of the GDS (Figure S1) reached this same position on June 8-9, mixing and with the local storm and engulfing it. Then the GDS expanded very rapidly in the zonal direction along the southern border of the polar cap at 50°S-60°S (Figure 1l and S1).

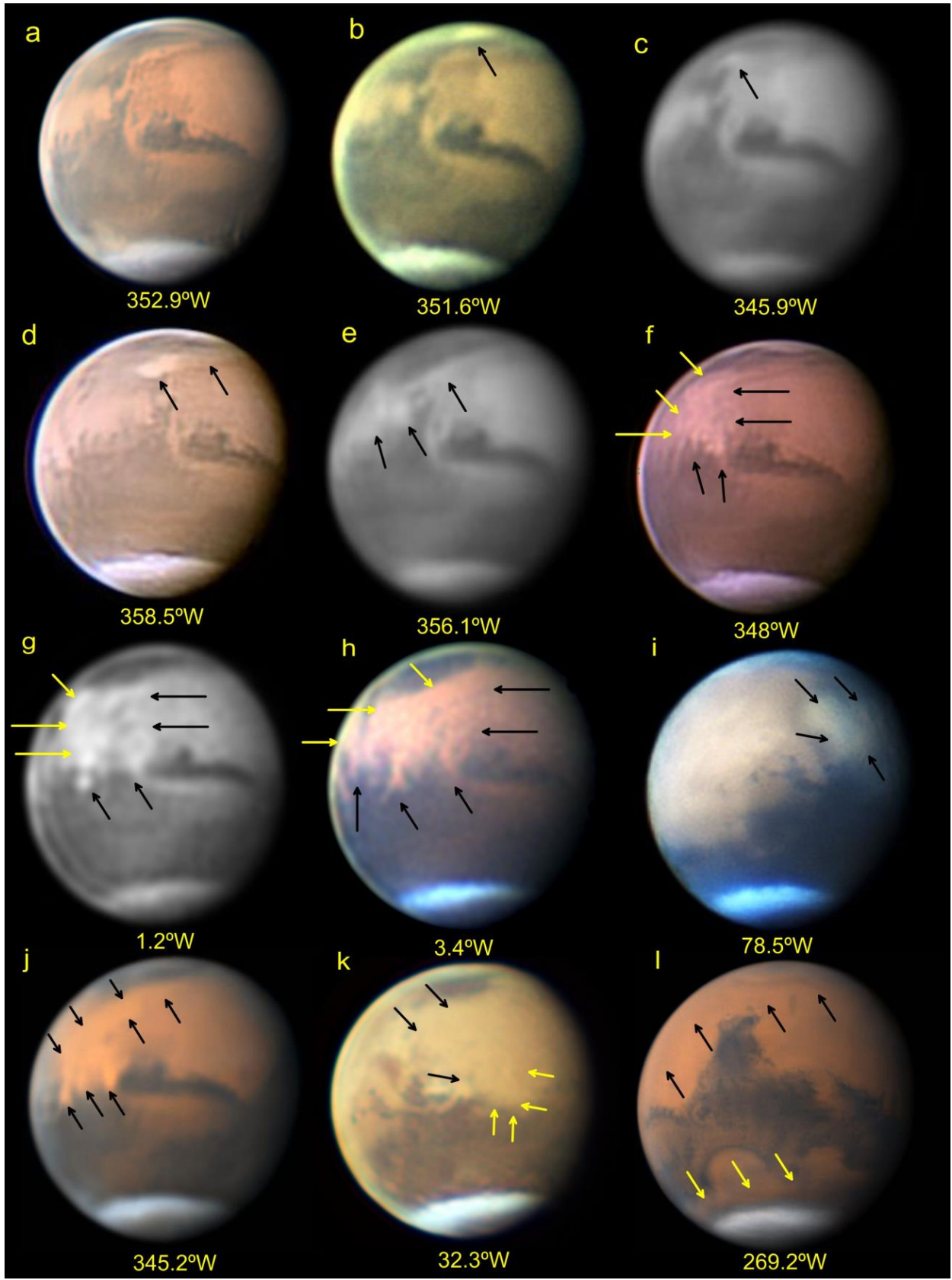

**Figure 1. The onset of the Global Dust Storm 2018.** *Image (a) shows the region free of storms (12 May, 20:51 UT, T. Olivetti) and (b) shows a precursor storm (arrow) centered at 346.7°W and 56°N (27 May, 05:42 UT, G. Grassmann). Images (c)-(l) show the onset and initial daily expansion phase of the GDS2018 (the arrows identify the area of the expanding storm). (c) 30 May, 7:25 UT, E. Morales; (d) 31 May, 08:55 UT, J. Rueck; (e) 1 June, 09:24 UT, E. Morales; (f) 2 June, 09:29 UT, J. Rueck; (g) 3 June, 11:02 UT, P. Maxon; (h) 4 June, 11:49 UT, K. Beverage; (i) 4 June, 16:57.9 UT, D. Millika & Nicholas; (j) 5 June, 11:13.4 UT, D. Peach; (k) 6 June, 15:05 UT, A. Casely; (l) 8 June, 07:55.9 UT,D. Peach. The Central Meridian (CM) is given in the west longitude system below each image.*

We tracked the motion of the edges of the dusty area from May 31 to June 8 in those regions where the dust showed high contrast with surface albedo marks (because of its high optical depth) (Figure 2a-b and Table 1). From the drift rate of these features (measured in longitude and latitude degrees per unit time) we retrieve the zonal ($U$) and meridional ($V$) velocities (with positive $U$ and $V$ indicating eastward and northward motion respectively). We found the following results for the meridional expansion:

(1) There was an initial northeastward displacement of the dust between 30 to 31 May from the storm area at 35°N to 45°N, implying velocities $U = 17 \pm 5$ ms$^{-1}$.

(2) We did not observe transport of dust poleward of 47°N (with $V$ in the range from - 0.2 to + 0.1 ms$^{-1}$).

(3) The zonal expansion along the equatorial area took place with mean velocities U = - 6 $\pm$ 2 ms$^{-1}$ (westward) and + 5.6 $\pm$ 2 ms$^{-1}$ (eastward).

(4) The southward velocity of the dust from 20°N to 40°S along the meridional branches was $V = - 5 \pm 1$ ms$^{-1}$. Afterwards, dust moving with $V = - 14 \pm 4$ ms$^{-1}$ reached latitude 60°S, close to the edge of the South polar region.

(5) On June 7-8, when the dust reached the south polar cap edge at 60°S, a rapid eastward expansion was noted with $U = 40 \pm 5$ ms$^{-1}$.

In Figure 2c, we show the growth of the area of the storm during this period and compare it with previous cases where this information is available. In the case of the 2001 GDS, we used the curve published by Cantor (2007). For the 1977A and B GDSs we retrieved the areas from published contour maps of the dusty region by Thorpe (1979) and in the case of the 2007 GDS we measured the area in the available MRO/MARCI maps by H. Wang (Wang and Richardson, 2015). In the case of the 2018 GDS, we have considered two separate regions: what we call the core, which we identify with the region with higher reflectivity and opacity with respect to the background albedo (i.e. where higher optical depths occur, showing well-defined borders), and the total area that includes the surrounding region to this core, with lower optical depth. Details on the measurement method are given in the Supplementary Information.

The growing rate of the area of the GDSs was not linear, as it is apparent in Fig. 2c. After ~ 6 days of expansion, the area of the 2018 GDS surpassed the critical value of $10^7$ km$^2$ proposed by Cantor (2007) as the maximum size for a storm to be considered as a regional storm (Wang and Richardson, 2015). In order to characterize and compare the growing rates of the confirmed GDS for which we have the area expansion, we have fitted the expansion curves in Fig. 2c to second order polynomials:

GDS 2001: $A(t) = 2.62 - 2.16t + 3.35t^2$

GDS 2007: $A(t) = 2.33 - 3.39t + 5.04t^2$

GDS 2018: $A(t) = 2.30 - 0.42t + 3.09t^2$

where the area $A$ is measured in $10^5 km^2$ and time $t$ in days, with $t$=0 corresponding to the first detection of the storm. For the 2018 GDS, the initial expansion rate (1-3 June) was dA/dt ~ $1.5 \times 10^6$ $km^2 day^{-1}$ and increased in 5-8 June to dA/dt ~ $4.3 \times 10^6$ $km^2 day^{-1}$. This number is similar to that measured during the 2001 GDS (dA/dt ~ $5 \times 10^6$ $km^2 day^{-1}$) but lower than in the GDS 2007 storm, for which we find dA/dt ~ $6.7 \times 10^6$ $km^2 day^{-1}$.

In Figure 3 (upper panel) we show a map of Mars with the daily evolution of the area covered by the 2018 storm during the period under analysis. Note the southward expansion of the dust from the onset area, accompanied by its east-west progression. Note also the northern confinement of the dust up to 45°N, whereas in the south it reached 60°S and then was rapidly transported eastwards at the edge of the south polar cap. After June 8, tracking the expansion of the storm became more difficult due to the low contrast in the border between dusty and non-dusty regions. However, the strong decrease in the visibility of the surface albedo features was evident throughout the month of June, when all longitudes were covered by the dust as shown in Fig. S2. As reported by Guzewich et al., 2019, the eastward progression of the GDS produced a first increase in dust opacity at the Curiosity location (4.6°S, 222.6°W) on 9-10 June reaching its peak in opacity in June 18[th]. On June 18, we see that the dust had covered the Tharsis Mons area (~ 15°N, 140°W), and most of the surface albedo features of this region and surrounding areas were unrecognizable on June 27. Likewise, dust continued to expand towards the South Pole, penetrating the polar cap by mid-June (Hernandez-Bernal. et al., in preparation 2019). Simultaneous to the GDS progression, secondary local storms (length scale ~ 500-1,000 km) occurred from 10 to 19 June westwards of the GDS (80°W-130°W and 20°S-30°S) as shown in Fig. S3. Other local storm events took place within the GDS region, in Syrtis Major (~ 20°N, 295°W) on June 23 (Fig. S2).

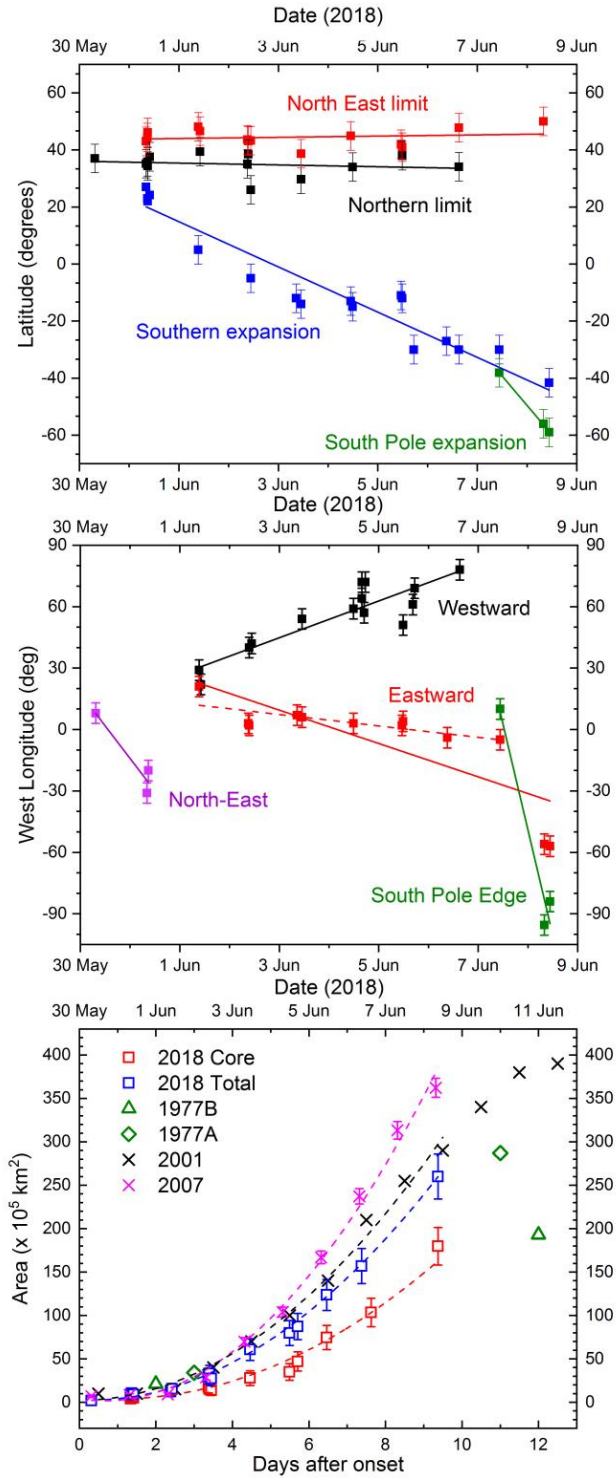

**Figure 2. Motions and area expansion of the dust.** Latitude (top) and west longitude (middle) drift rates of selected features at the borders of the GDS area from 30 May to June 8. Linear fits

have been performed to retrieve mean velocities along this period. The bottom panel shows the evolution of the total area occupied by the dust (blue squares) and that of the "core" (central regions of the GDS with higher optical depth) (red squares). For comparison, we show the areas of the GDS 2001 (from Cantor, 2007), estimations for the GDS 1977A and B based on maps published in Thorpe (1979) and measurements from available MRO/MARCI maps of the 2007 case (Wang and Richardson, 2015).

Table 1

GDS 2018 motions and velocities

| Feature | Dates | Latitude range | Longitude range | Drift (°/day) | Velocity (ms⁻¹) |
|---|---|---|---|---|---|
| Westward border | 1-6 June | 7°S to 20°N | - | +9.0 | U = -6.2 ± 2 |
| Eastward border | 1-8 June | 8°S to 10°N | - | -8.1 | U = 5.6 ± 2 |
| North-East limit | 30-31 May | 35°N to 45°N | - | -31.8 | U = 17 ± 4 |
| South Pole edge | 7-8 June | 55°S | - | -103 | U = 40 ± 5 |
| North-East limit | 31 May- 8 June | - | 330°W-360°W | +0.2 | V = +0.15 ± 1 |
| Northern limit | 30 May – 6 June | - | 5°W - 25°W | -0.3 | V = -0.2 ± 1 |
| Southern border | 31 May – 8 June | - | 10°W-30°W | -7.3 | V = -5 ± 1 |
| South Pole border | 7-8 June | - | 10°W | -20.5 | V = -14 ±4 |

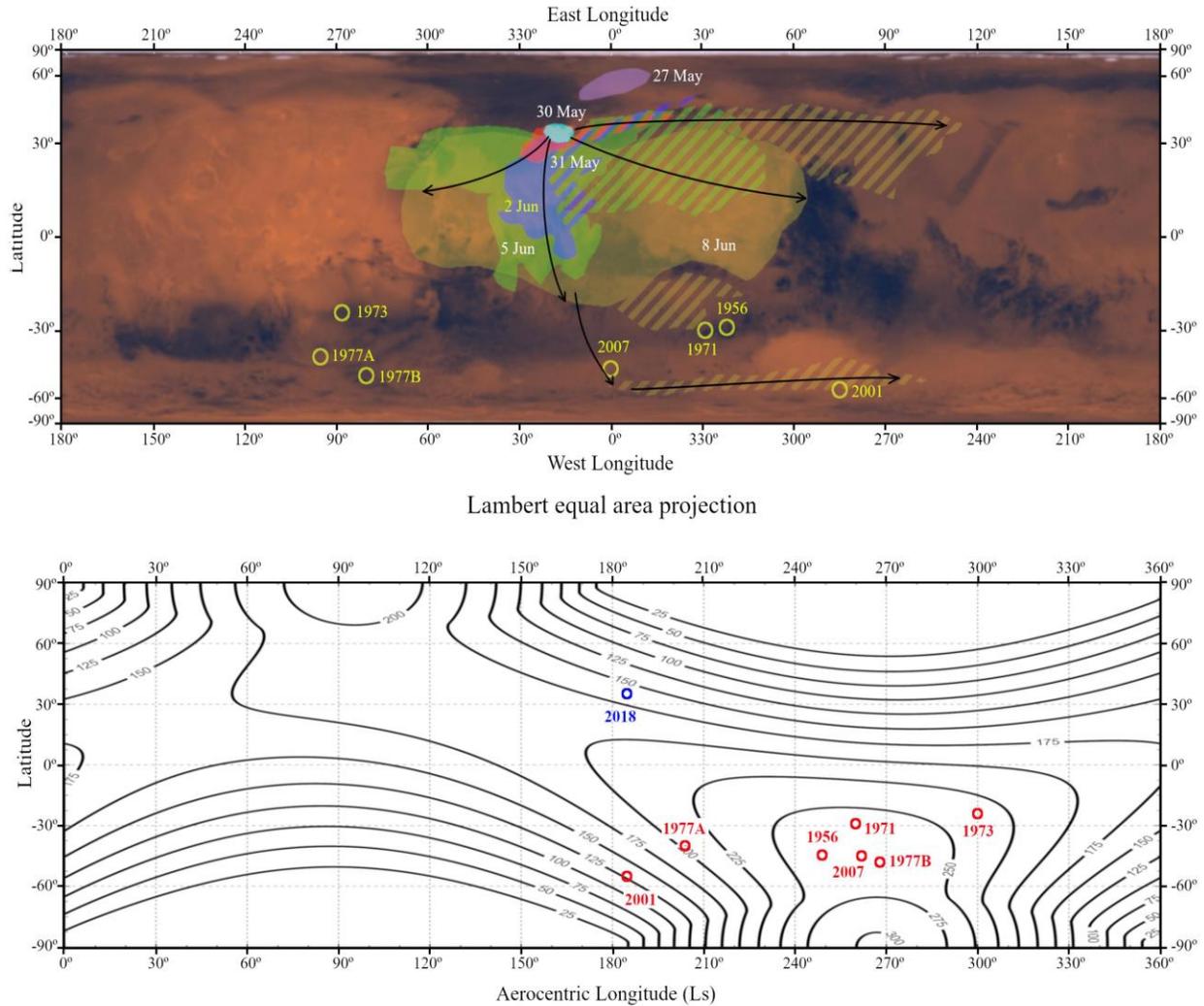

**Figure 3.** **Maps of the 2018 GDS expansion and GDS occurrence.** Upper: Lambert equal-area projection of Mars, showing the expanding area of the storm from 30 May to 8 June superimposed with different colors. Flat colors indicate the core of the storm, and streaked regions the total area. The purple area on 27 May corresponds to a precursor storm. Arrows indicate the directions followed by the expanding dust. Lower: Mean daily insolation (Wm$^{-2}$) at the top of the atmosphere along a Martian year (adapted from Sánchez-Lavega et al., 2018). The circles in both panels mark the onset location of the confirmed GDSs given in Table S1.

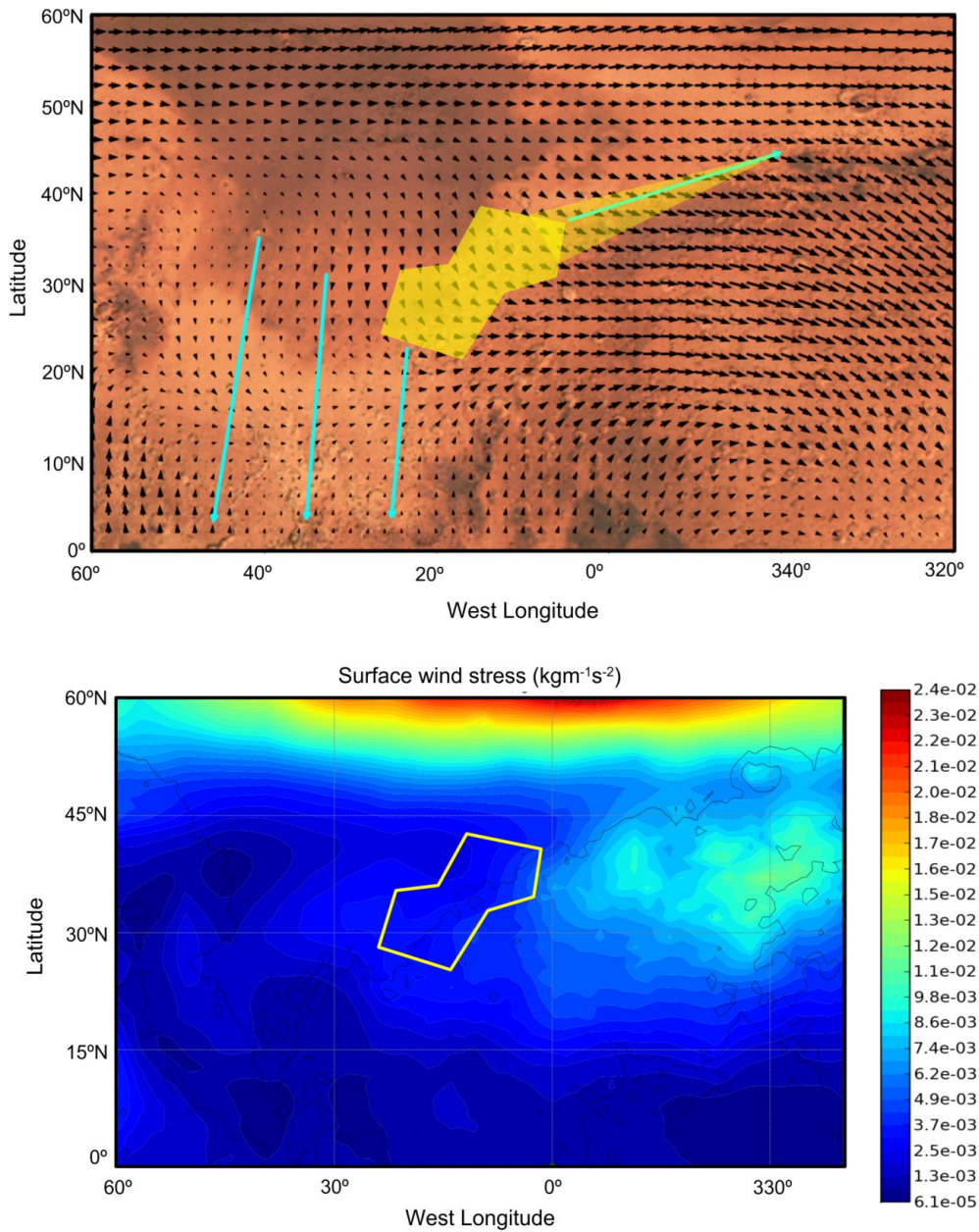

**Figure 4. Model predictions at the GDS 2018 onset epoch.** The maps show, on a surface albedo map, the statistical predictions of the Mars Climate Database (MCD) using the climatology scenario (Forget et al., 1999; Millour et al., 2015; http://www-mars.lmd.jussieu.fr/mcd_python/). Dark arrows in the upper map show surface wind prediction for 31 May 2018 at 8:14 U.T., corresponding to the observed onset time (Fig. 1 and Table S1). The maximum wind speeds are 15 ms$^{-1}$. The yellow polygon corresponds to the approximate area covered by the storm and the light triangle to the initial Northeast expansion. Blue arrows mark the observed GDS initial expansion directions. The lower map shows the surface-wind stress predicted at the same time.

### 3. Discussion on the onset of the 2018 GDS

Figure 3 shows the location of the onset of the confirmed GDS events as given in Table S1, both on a map of Mars and on chart showing the mean daily seasonal insolation as a function of latitude and solar longitude. Both representations show clearly that the onset of the 2018 event was different to previous cases. On the one hand, it occurred for the first time in northern latitudes. On the other, it appeared at the earliest time of the Martian year (Ls = 184°), the same as the 2001 event (Strausberg et al., 2005; Cantor, 2007). The onset occurred in the region of Acidalia Planitia, known to be a Martian region where many regional dust storms initiate (Wang and Richardson, 2015). However, this is the first reported case where a storm developing in Acidalia evolves to a GDS event. The growing pattern followed by the GDS 2018 fits very well the evolution of storms in this region, as described by Wang and Richardson (2015). The storm expanded north to south, crossing the equator along a known route of progression, the low-topographic corridor of Acidalia (compare Fig.3 upper map with Fig. 6 in Wang and Richardson, 2015). This was accompanied by simultaneous east and west expansions (Fig 3, arrows in upper map). West of the Acidalia route, the development style was a sequential activation of storm branches oriented from north to south, followed by their mutual merger, in a similar behavior described for previous storms (Wang, 2007; Wang and Richardson, 2015). Fig. 3 also shows that both the 2001 and 2018 events occurred under low daily insolation conditions, corresponding to the northern autumnal equinox (insolation at the Top of the Atmosphere (TOA) ~ 125 - 150 Wm$^{-2}$). This value is lower than that of the rest of GDS cases, which took place in southern summer solstice, with TOA values ~ 200-260 Wm$^{-2}$. During the dusty (perihelion) season (Ls = 180° to 360°), Acidalia remains most of the time under low insolation (TOA < 125 Wm$^{-2}$) except during the narrow period from Ls ~ 180° to 230°. Since radiative heating by the dust is a well known ingredient for generating a GDS, this could represent the preferred period in Acidalia where a GDS can initiate, and might be a reason for the rarity of GDS starting in the northern hemisphere. Northern baroclinic waves might also play a significant role in the case of the onset of the 2018 event, as they do for other storms initiating in the "storm zones" and developing along the "storm tracks" of Acidalia, Utopia and Arcadia (Montabone, L. and Forget, 2018b).

In addition to the topographic influence and dust radiative heating, another ingredient of the recipe to trigger a GDS is an appropriate coupling with the atmosphere general circulation at the onset season (Zurek et al., 1992; Barnes et al., 2017; Khare et al., 2017). For the most common GDSs, occurring in the south latitudes during the southern summer solstice (Fig. 3), it has been proposed that their explosive expansion is the result of a radiative-dynamical coupling with the Hadley circulation. This coupling results in a positive feedback between the increase of dust heating and the alteration of the Hadley cell structure followed by enhanced convergence and dust motion and spread (Khare et al., 2017 and references therein). However, this mechanism cannot explain the 2001 and the 2018 global events, since in their near equinoctial period (autumn in the northern hemisphere for 2018) the Hadley cell is symmetric about the equator where convergence occurs, according to the MCD (Forget et al., 1999; Millour et al, 2015). Figure 4 shows that the direction and intensity of the winds do not correspond to the storm's expansion movements at the onset phase. Moreover, the MCD predictions of intensity of the surface wind stress do not correspond with the generation of a storm in that area at this time of year. The observed south propagation of the 2018 event probably requires intensification of the northern branch of the Hadley cell and a simultaneous suppression of the southern branch, since its mean circulation is equatorward, contrary to the motion observed in the GDS. As an alternative to this mechanism, thermal tides were proposed to have played a prominent role in the expansion of the 2001 GDS, which occurred at the same epoch than the 2018 event (Wilson, 2012). A transient teleconnection effect was also invoked for the 2001 case (Montabone et al., 2008; Martinez-Alvarado et al., 2009). The interference between two modes of the thermal tides (i.e. the westward-propagating diurnal mode and the eastward propagating diurnal one, equivalent to a "Kelvin mode") may induce a change in surface wind stress patterns in regions distant from the initial dust outburst, triggering secondary dust lifting centers and storm expansion. The new 2018 event adds together with the 2001 case, a challenge to the understanding of onset and expansion of "planet-encircling" dust storms.

## Acknowledgments, Samples, and Data

This work has been supported by the Spanish project AYA2015-65041-P (MINECO/FEDER, UE) and Grupos Gobierno Vasco IT-765-13. JHB was supported by ESA Contract No. 4000118461/16/ES/JD, Scientific Support for Mars Express Visual Monitoring Camera. We thank Luca Montabone and an anonymous reviewer for their constructive comments on the paper.

All the images and data used in this study can be accessed through the ALPO-Japan and PVOL2 databases. A list of the sources for the images used in this paper can be found in the Supporting Information.

# The Onset and growth of the 2018 Martian Global Dust Storm

A. Sánchez-Lavega[1,2*], T. del Río-Gaztelurrutia[1,2], J. Hernández-Bernal[1,2], M. Delcroix[3]

1. Dpto. Física Aplicada I, EIB, Universidad País Vasco UPV/EHU, Bilbao, Spain

2. Aula EspaZio Gela, Escuela de Ingeniería de Bilbao, Universidad del País Vasco UPV/EHU, Bilbao, Spain

3. Societé Astronomique de France, Paris, France.

* To whom correspondence should be addressed at: agustin.sanchez@ehu.eus

**Contents of this file**

## Introduction

This Supporting Information includes the following material. (1) A list of the contributors to the different image databases whose images have been used in this study and links to the databases where these images can be found; (2) A detailed description of the measurements, procedures and analysis we have followed for the different kind of images.

### Text S1
**Data availability**
This work relies on images that can be downloaded from the following sources:
(1) Association of Lunar and Planetary Observers ALPO – Japan:
http://alpo-j.asahikawa-med.ac.jp/Latest/Mars.html
(2) PVOL2 database: http://pvol2.ehu.eus/pvol2/
(3) Internet networks and forums surveyed by M. Delcroix (with authors' permissions).
Data available from M. Delcroix (delcroix.marc@free.fr) upon request.
Observers details are given in Table S1

### Text S2
**Image analysis**
Ground-based images used in this study were obtained employing the "lucky imaging" method (Mousis et al., 2014). Most telescopes employed were in the range 0.3-0.5 m in diameter (Table S2). At the time of the GDS onset, Mars subtended 15 arcsec on the sky and for a typically used telescope of 36 cm in aperture it implies a resolution of ~ 175 km at Equator. D. Peach contributed with a set of images obtained using "Chilescope" (http://www.chilescope.com/), a remotely controlled 1 m telescope. The images span the spectral ranges ~ 450-650 nm (from color composites Red-Green-Blue, RGB) and the near infrared (~ 685 nm-980 nm). The list of contributors whose images have been used in this study is given in the Table S2. Images were navigated to fix the Martian disk using the WinJupos free software (2019) and in most cases reprocessed to increase the contrast of weak features. Navigation errors (limb location and longitude-latitude coordinate grid fixing over the disk of Mars, and cursor pointing) translates in longitude errors of 3º at high latitudes and 1º at Equator, and latitude errors of 1º-2º.

To determine the area covered by the storm, we generated Lambert equal-area maps from navigated images using WinJupos (http://www.grischa-hahn.homepage.t-online.de/, accessed 2019). In our projections, a pixel corresponds to an area of 278 km². We used the freely available software GIMP (https://www.gimp.org) to select both kinds of regions in the projections, and to count the number of pixels within the selected regions. Error was estimated using the formula $\Delta A = 2\sqrt{A}\Delta L$, where $\Delta L$ is the linear incertitude in the location of the border of the region, that we took to be 15 pixel, corresponding to 250km.

**Table S1**

**Confirmed Martian Global Dust Storms** (*)

| Year | Martian Year (MY) | Ls | Duration | Initiation location | Notes |
|------|-------------------|------|-------------|----------------------|-----------|
| 1956.63 | 1 | 249° | Aug 19-Nov | 31°E, 30°S | (1) |
| 1971.72 | 9 | 260° | Sep 22-Jan | 38°E, 29°S | (2) |
| 1973.81 | 10 | 300° | Oct 13-Dec | 272°E, 24°S | (3) |
| 1977.12 | 12 | 204° | Feb 15-April | 265°E, 40°S | (4) |
| 1977.23 | 12 | 268° | May 27-Oct | 280°E, 48°S | (5) |
| 2001.48 | 25 | 185° | June 26-Oct | 75°E, 55°S | (6) |
| 2007.47 | 28 | 262° | June 22-Oct | 0°E, 45°S (**) | (7) |
| 2018.41 | 34 | 184.9° | May 30-Aug | 348°E, 35°N | This work |

<u>Notes:</u> Global Dust Storms (GDS) or "planet encircling"

(*) Following Khare et al. (2018), adding the current case (this work).
(1) Telescopic and photographic. Martin and Zurek (1973)
(2) Telescopic and photographic. Mariner 9; Martin and Zurek (1973)
(3) Telescopic and photographic. Martin and Zurek (1973)
(4) Viking lander and orbiter. Thorpe (1979; 1981); Ryan and Sharman (1981); Khare et al. (2017).
(5) Viking lander and orbiter. Thorpe (1979; 1981); Ryan and Sharman (1981); Khare et al. (2017).
(6) Seven precursor storms. Telescopic; Mars Global Surveyor; Hubble Space Telescope. Cantor (2007), Strausberg et al. (2005); Khare et al. (2018).
(7) (**) Ambiguity exists on its initial source. A "precursor storm" occurred in the North (Chryse Planitia, ~ Equator, 330°E). However, the explosive storm leading to the GDS occurred in Noachis (the position given in the Table). Telescopic; Mars Reconnaissance Orbiter. See details in Wang and Richardson (2015).

**Table S2**

**Mars Global Dust Storm: outbreak & early stages**
Selected observations from 12 May – 8 June (2018)

| Observer | Country | Instrument | Filters | Days | GDS |
|---|---|---|---|---|---|
| Tiziano Olivetti | Thailand | DK 505mm | R,G,B | 12May - 20:51 | N (survey) |
| Alfredo Vidal | Spain | SC360mm | IR714 | 19May - 01:47 | N(survey) |
| Guilherme Grassmann | | Newton 12" 305mm | R,G,B | 27May - 05:15, 05:52 | N - Precursor |
| Fabio Carvalho | Brazil | Newton 406mm | R,G,B, IR | 27 May – 06:54.0 | N - Precursor |
| Eric Sussenbach | Curaçao | SC200mm | R,G,B | 27May - 07:55 | N - Precursor |
| Efrain Morales | Puerto Rico | SC 305mm | R,G,B,IR685 | 30 May - 7:25 31May - 7:31 | Y - Onset |
| Jean Boudreau | USA | DK368mm | R,G,B | 31 May – 08:52 | Y |
| Joseph Rueck | USA | Newton 460mm | R,G,B | 31May - 08:55 – 09:54 | Y |
| Paul Maxon | USA | Mewlon 250mm | R,G,B,IR742 | 31May - 10:49 | Y |
| Efrain Morales | Puerto Rico | SC305mm | IR685 | 1 June - 09:24 | Y |
| Tayler Schubauer | USA | SC203mm | Color | 1 June 10:10.0 | Y |
| Paul Maxon | USA | Mewlon 250mm | R,G,B, IR742 | 1 June - 10:46 | Y |
| John Kazanas | Australia | Newton 320mm | R,G,B | 1 June - 16:26 | N (at 99°W) |
| Teruaki Kumamori | Japan | SC 360 mm | L,R,G,B | 1 June - 16:40 | N (at 100°W) |
| Yaroslav Naryzhniy | Ukraine | Newton 320mm | R,G,B | 2 June 01:27.0 | Y |
| Joseph Rueck | USA | Newton 460mm | R,G,B | 2 June – 09:29.3 | Y |
| Paul Maxon | USA | Mewlon 250mm | R,G,B,IR742 | 2 June – 10:47 | Y |
| Milika & Nicholas | Australia | SC360mm | L,R,G,B | 2 June – 17:22 | Y (partial) |
| Amrit Seecharan | Trinidad | SC235mm | Color | 3 June | Y |
| Paul Maxon | USA | Mewlon 250mm | R,G,B,IR742 | 3 June – 11:02 | Y |
| Paul Maxon | USA | Mewlon 250mm | R,G,B,IR752 | 4 June – 10:26 | Y |
| Kevin Beverage | USA | SC254mm | R,G,B | 4 June - 11:49.4 | Y |
| R. Iwamasa | Japan | SC360 mm | L,R,G,B,IR685 | 4 June – 15:54 | Y (partial) |

| Tsuyoshi Arakawa | Japan | New 300mm | R,G,B | 4 June -15:52.1 | Y (partial) |
|---|---|---|---|---|---|
| S. Ota | Japan | New 300mm | R,G,B | 4 June – 17:36.7 | Y (partial) |
| Milika & Nicholas | Australia | SC360mm | L,R,G,B | 4 June – 16:57.9 | Y (partial) |
| Akihiro Yamazaki | Japan | C400mm | L,R,G,B | 4 June – 17:29.1 | Y (partial) |
| Damian Peach | UK-Chile | T1000mm | R,G,B | 5 June – 07:22-11:13 | Y |
| Paul Maxon | USA | Mewlon 250mm | R,G,B,IR742 | 5 June – 06:23 | Y(partial) |
| Kevin Beverage | USA | SC254mm | R,G,B | 5 June - 11:43.5 | Y |
| Milika-Nicholas | Australia | SC360mm | L,R,G,B | 5 June – 16:23-18:29 | Y (partial) |
| John Kazanas | Australia | Newton 320mm | R,G,B | 5 June – 17:17 | Y (partial) |
| Isamu Hirabayashi | Japan | Newton 250mm | R,G,B | 4 June – 16:34 | Y (partial) |
| Steven Yockey | USA | SC280mm | R,G,B | 6 June - 09:03.0 | Y |
| Paul Maxon | USA | Mewlon 250mm | R,G,B, IR742 | 6 June – 10:16.1 | Y (partial) |
| Andy Casely | Australia | SC360mm | R,G,B,IR642 | 6 June – 15.05.0 | Y |
| Paul Maxon | USA | Mewlon 250mm | R,G,B, IR742 | 7 June – 10:17.1 | Y (partial) |
| Isamu Hirabayashi | Japan | Newton 250mm | R,G,B | 7 June – 16:36.3 | Y (partial) |
| Damian Peach | UK-Chile | T1000mm | R,G,B | 8 June – 07:55.9 | Y (partial) |
| Paul Maxon | USA | Mewlon 250mm | R,G,B, IR742 | 8 June – 10:19.9 | Y (partial) |

Note.
The Table gives the following information. Observer name and country, telescope type and objective (aperture) in mm, filters used (standard RGB filters: R for red, G for green, B for blue, L for luminance), infrared cutting filters (IR 642, IR685, IR714, IR742, IR752 transmission above the wavelength given in nm), observing date and time, and GDS coverage (Y = yes, N= no, partial).

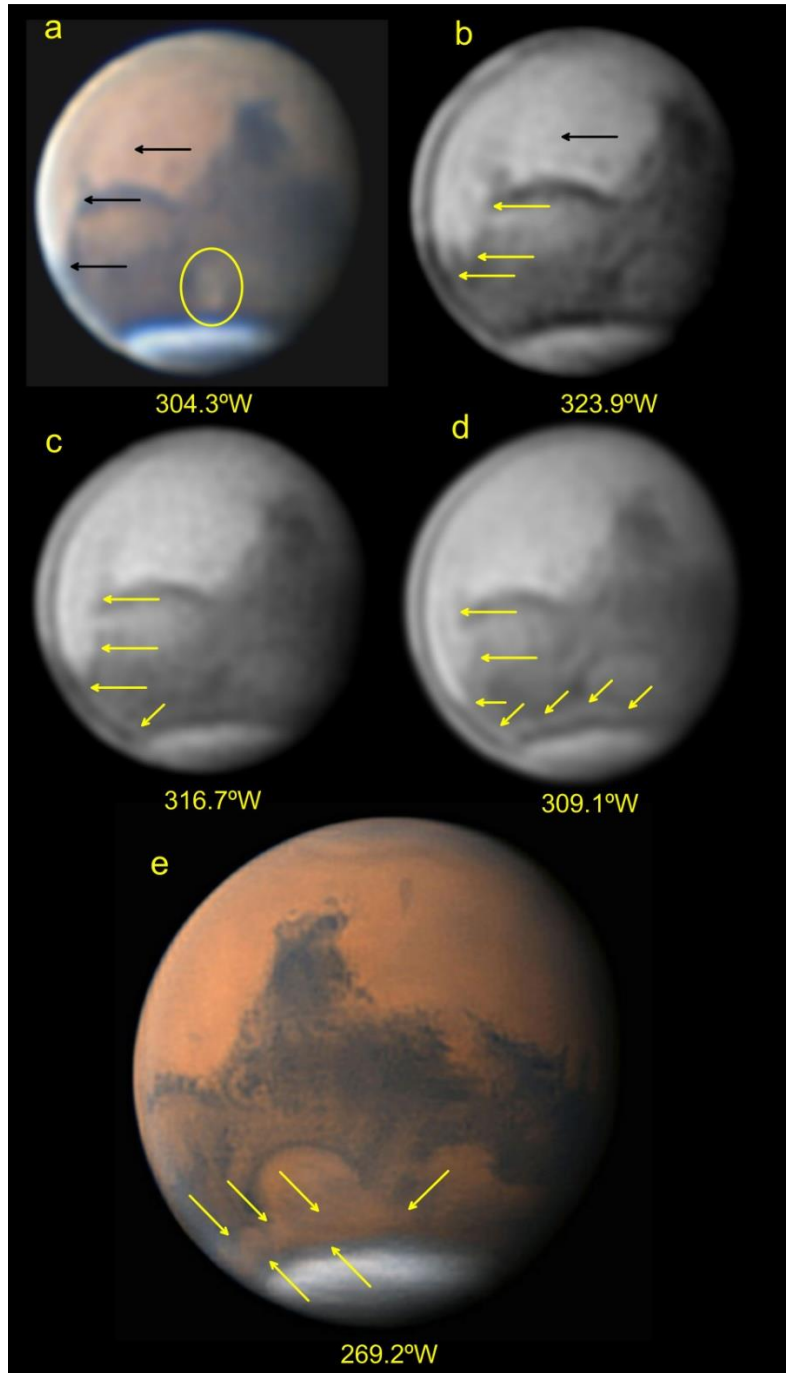

**Figure S1.** Images showing the southern hemisphere expansion of the GDS 2018 from 6 to 8 June. (a) 6 June, 09:03.5 UT, S. Yockey; (b) 6 June, 10:24 UT, P. Maxon (742 nm); (c) 7 June, 10:33 UT, P. Maxon (742 nm); (d) 8 June, 10:40 UT, P. Maxon (742 nm); (e) 8 June, 07:55.9, D. Peach. In (a) the oval marks the location of a local storm in the western side of Hellas. The Central Meridian (CM) is given in the west longitude system below each image.

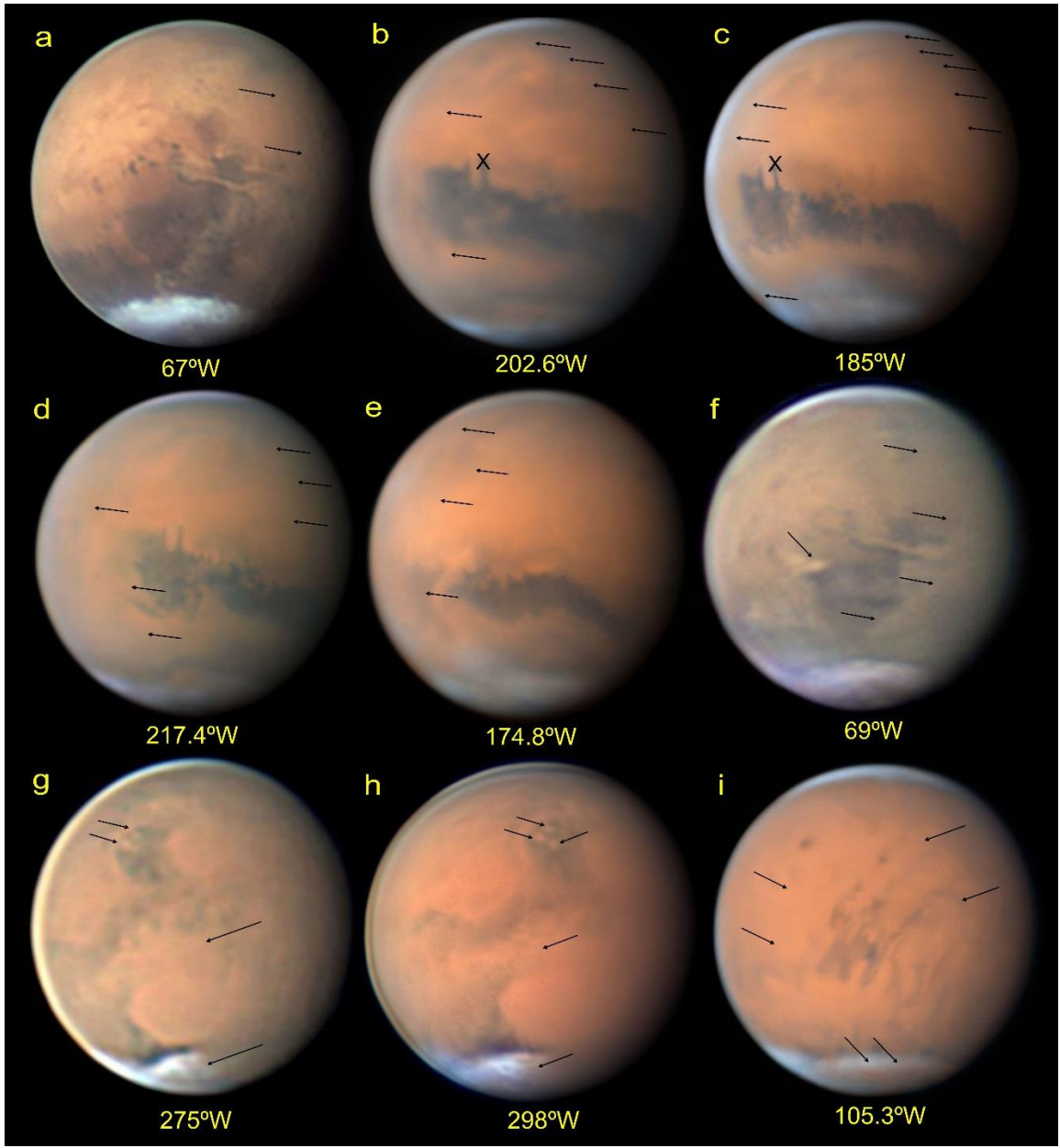

**Figure S2.** Images showing the evolution of the GDS 2018 from 9 to 27 June 2018. (a) 9 June, 19:25.6 UT, A. Wesley; (b) 15 June, 07:49.7 UT, D. Peach; (c) 16 June, 07:15.5 UT, D. Peach; (d) 17 June, 10:06.5 UT, D. Peach; (e) 18 June, 07:49.8 UT, D. Peach; (f) 19 June, 01:13.7 UT, C. Foster; (g) 23 June, 17:49.4 UT, A. Wesley; (h) 23 June, 19:23 UT, , C. Go; (i) 27 June, 08:42.9 UT, D. Peach. The Central Meridian (CM) is given in the west longitude system below each image. Crosses in (b) and (c) mark the location of crater Gale and MSL-Curiosity.

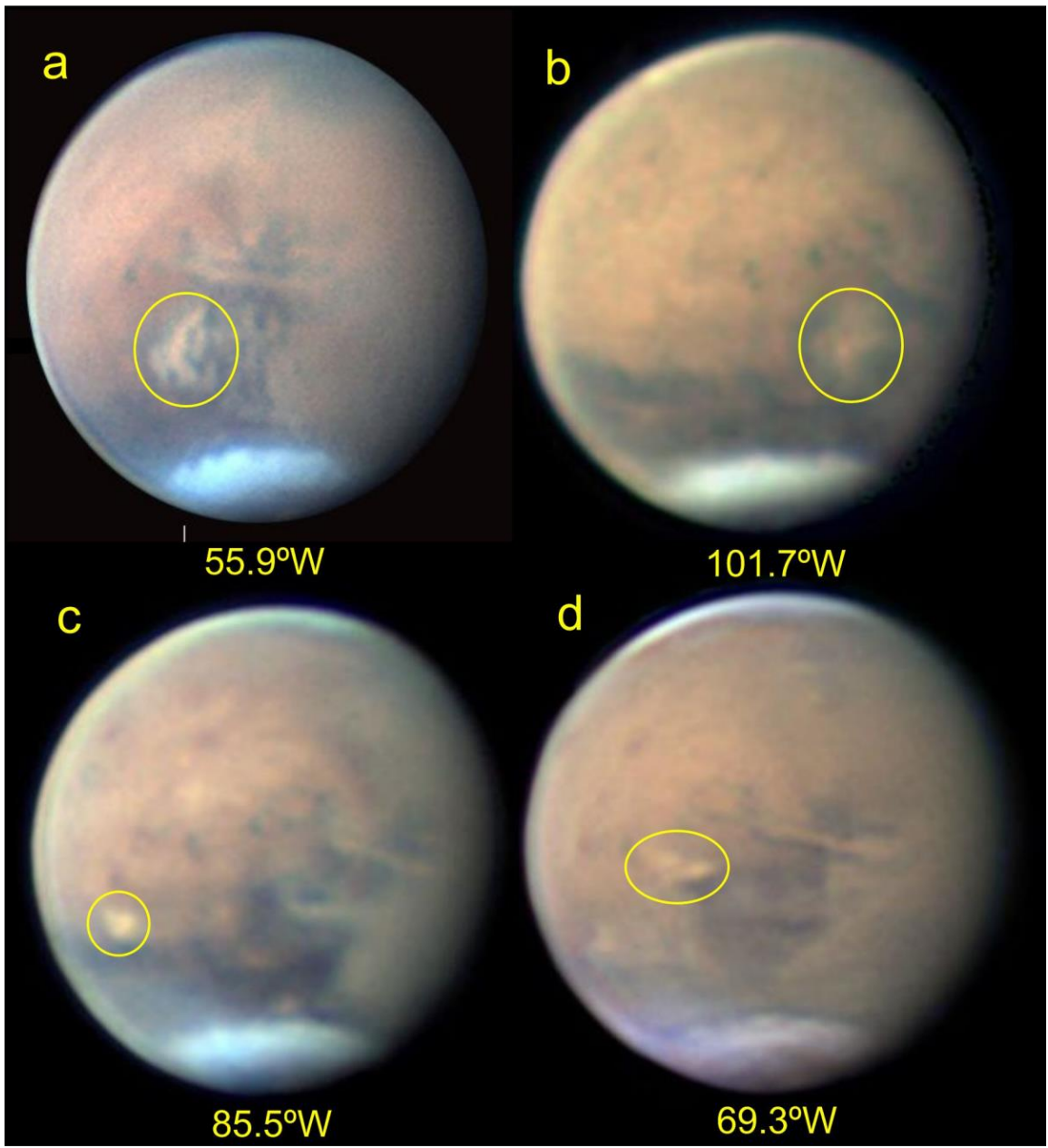

**Figure S3.** Secondary storms developing during the GDS 2018 identified within yellow circles. (a) 10 June, 19:15 UT, S. Ota; (b) 10 June, 22:23 UT, C. Foster; (c) 12 June, 22:33 UT, C. Foster; (d) 19 June, 01:14 UT, C. Foster. The Central Meridian (CM) is given in the west longitude system below each image.